\documentclass{nature}

\usepackage{amsmath}    % need for subequations
\usepackage{color}
\usepackage[10pt]{moresize}

\usepackage{graphicx}
\usepackage{caption}
\DeclareCaptionLabelFormat{adja-page}{\hrulefill\\#1 #2 \emph{(previous page)}}
\usepackage{subfig}

\usepackage{amsfonts}
\usepackage{amssymb}
\usepackage{amscd}
\usepackage{enumerate}
\usepackage{epsfig}
\usepackage{epstopdf}
\usepackage{dcolumn}% Align table columns on decimal point
\usepackage{bm}% bold math
\usepackage{amsthm}
\usepackage{amsfonts}
\usepackage{color} % Para incluir texto en rojo.
\usepackage[usenames,dvipsnames]{xcolor}
\usepackage{amsmath}
\usepackage{enumerate}
\usepackage{bbm}
\usepackage[colorlinks=true,citecolor=blue,urlcolor=black]{hyperref}

\newcommand{\ket}[1]{\mbox{$\left| #1 \right\rangle$}}

\begin{document}

\title{Experimental unconditionally secure bit commitment}

\author{Yang Liu$^{1,\ast}$, Yuan Cao$^{1,\ast}$, Marcos Curty$^{2,\ast}$, Sheng-Kai Liao$^{1}$, Jian Wang$^{1}$, Ke Cui$^{1}$, Yu-Huai Li$^{1}$, Ze-Hong Lin$^{1}$, Qi-Chao Sun$^{1}$, Dong-Dong Li$^{1}$, Hong-Fei Zhang$^{1}$, Yong Zhao$^{1,3}$, Cheng-Zhi Peng$^{1}$, Qiang Zhang$^{1}$, Ad\'an Cabello$^{4}$, Jian-Wei Pan$^{1}$}

\maketitle

\begin{affiliations}
\item Shanghai Branch, Hefei National Laboratory for Physical Sciences at Microscale and
Department of Modern Physics, University of Science and Technology
of China, Hefei, Anhui 230026, P.~R.~China.
\item Department of Signal Theory and Communications, University of Vigo, E-36310 Vigo, Spain.
\item Shandong Institute of Quantum Science and Technology Co., Ltd, Jinan, Shandong 250101, P.~R.~China.
\item Departamento de F\'{\i}sica Aplicada II, Universidad de Sevilla, E-41012 Sevilla, Spain.\\
$^\ast$ These authors contributed equally to the paper.
\end{affiliations}

\baselineskip24pt

\maketitle

%%%%%%%%%%%%%%%%%%%%%%%%%%%%%%%%%%%%%%%%%%%%%%%%%%%%%%%%%%%%%%%%%%%

\begin{abstract}
Bit commitment is a fundamental cryptographic task that guarantees a secure commitment between two mutually mistrustful parties and is a building block for many cryptographic primitives, including coin tossing~\cite{Blum82,BRA00}, zero-knowledge proofs~\cite{GOLD85,GOLD86}, oblivious transfer~\cite{BEN91,UN10} and secure two-party computation~\cite{KILIAN88}. Unconditionally secure bit commitment was thought to be impossible~\cite{Mayers96,Mayers97,LC96,LC97,MKP04,DKSW07} until recent theoretical protocols that combine quantum mechanics and relativity were shown to elude previous impossibility proofs~\cite{Kent99,Kent05,Kent11,Kent12}.
Here we implement such a bit commitment protocol~\cite{Kent12}. In the experiment, the committer performs quantum measurements using two quantum key distribution systems~\cite{BB84} and the results are transmitted via free-space optical communication to two agents separated with more than $20$ km. The security of the protocol relies on the properties of quantum information and relativity theory. We show that, in each run of the experiment, a bit is successfully committed with less than $5.68\times10^{-2}$ cheating probability. Our result demonstrates unconditionally secure bit commitment and the experimental feasibility of relativistic quantum communication.
\end{abstract}

%%%%%%%%%%%%%%%%%%%%%%%%%%%%%%%%%%%%%%%%%%%%%%%%%%%%%%%%%%%%%%%%%%%

Bit commitment is a cryptographic protocol between two distrustful parties. It has two phases. In the first (commit), the committer, Alice, carries out actions that commit her to a particular bit value $b$. In the second (reveal), Alice, if she so chooses, gives the receiver, Bob, information that unveils $b$. Bit commitment must be concealing and binding. It is
concealing if Bob cannot learn $b$ before Alice unveils it, and it is
binding if
Alice cannot change $b$ once she has committed to it.

In classical cryptography, bit commitment is achieved by utilising computational complexity assumptions such as, for instance, the difficulty of factoring large numbers.
However, the security of such schemes can be broken using a quantum computer\cite{ShorAlgorithm_94}.
Indeed, it can be proven that unconditionally secure bit commitment is impossible using only classical resources.
The same holds true even if Alice and Bob are allowed to use quantum resources in a non-relativistic scenario~\cite{Mayers96,Mayers97,LC96,LC97,MKP04,DKSW07}.
For this reason, quantum bit commitment schemes rely on physical assumptions as,
for example, that the attacker's quantum memory is noisy~\cite{steph}.
Interestingly, the picture changes dramatically if we take into account the signalling constraints implied by the Minkowski causality in a relativistic context. Then, assuming that quantum mechanics is correct and that space-time is approximately Minkowskian, it has been shown that there are bit commitment protocols offering unconditional security~\cite{Kent11,Kent12,KTHW12}.

%%%%%%%%%%%%%%%%%%%%%%%%%%%%%%%%%%%%%%%%%%%%%%%%%%%%%%%%%%%%%%%%%%%
% Protocol
%%%%%%%%%%%%%%%%%%%%%%%%%%%%%%%%%%%%%%%%%%%%%%%%%%%%%%%%%%%%%%%%%%%

The protocol~\cite{Kent12} implemented in our experiment involves six parties: Alice and her agents $A_0$ and $A_1$, and Bob and his agents $B_0$ and $B_1$. They are distributed in three locations which are almost aligned, as illustrated in Fig.~\ref{diagram}a. The protocol has the following five steps:

\begin{enumerate}
\item The first step guarantees the security of the communications between Alice and her agents. For this purpose, Alice uses quantum key distribution~\cite{QKD1,QKD2} (or, alternatively, a trusted courier) to share two secret keys, $K_{A_0}$ and $K_{A_1}$, with $A_0$ and $A_1$.

\item The protocol itself starts when Bob sends Alice $N$ signals ({\it e.g.}, phase-randomised weak coherent pulses) prepared in either horizontal, vertical, diagonal or antidiagonal polarised states, which Bob selects independently and randomly for each signal.

\item To commit to the bit value $0$ ($1$), Alice measures all the incoming signals in the rectilinear (diagonal) polarisation basis. Then, she uses a public channel to notify Bob which signals she has detected. Also, she encrypts her measurement results with the one-time pad (OTP)~\cite{vernam} using the secret keys $K_{A_0}$ and $K_{A_1}$, and sends them to $A_0$ and $A_1$.

\item To unveil the commitment, agents $A_0$ and $A_1$ decrypt the measurement results received from Alice and send them to Bob's agents $B_0$ and $B_1$, respectively.

\item To verify the commitment, Bob compares the results submitted by $A_0$ and $A_1$. If they are different, Bob rejects the commitment. Otherwise, he estimates a lower bound, $n_{\rm rect}$, for the number of single photons sent in the rectilinear basis and detected by Alice. Likewise, Bob does the same with the signals he sent in the diagonal basis. 
Let $n_{\rm e, rect}$ ($n_{\rm e, diag}$) be the total number of errors in the rectilinear (diagonal) basis.
Only when both $n_{\rm rect}, n_{\rm diag}\geq{}N_{\rm tol}$ and $n_{\rm e, rect}\leq{}E_{\rm tol}N_{\rm tol}$ ($n_{\rm e, diag}\leq{}E_{\rm tol}N_{\rm tol}$) Bob accepts the commitment as 0 (1), for some prefixed parameters $N_{\rm tol}$ and $E_{\rm tol}$ previously agreed by Alice and Bob. 
\end{enumerate}

%%%%%%%%%%%%%%%%%%%%%%%%%%%%%%%%%%%%%%%%%%%%%%%%%%%%%%%%%%%%%%%%%%%
% Security
%%%%%%%%%%%%%%%%%%%%%%%%%%%%%%%%%%%%%%%%%%%%%%%%%%%%%%%%%%%%%%%%%%%

The protocol described above is perfectly concealing. This is so because the communication between Alice and her agents $A_0$ and $A_1$ is guaranteed by the OTP. Also, Bob's knowledge of Alice's detected events does not give him any information about her committed bit.
See Appendix~A for a discussion of the security of the protocol against a dishonest Bob. There
is also proven that the protocol is
binding. Indeed, it can be shown that Alice's cheating probability rapidly approaches zero when $N_{\rm tol}$ increases, given that $E_{\rm tol}$ is not too large. In our experiment, this results in a total cheating probability below $5.68\times10^{-2}$. Note, moreover, that this value comes from a very simple upper 
bound for the cheating probability, which may not be tight. In reality, therefore, the cheating probability may be significantly lower. 

\begin{figure}
\begin{center}
 \includegraphics[scale=0.54]{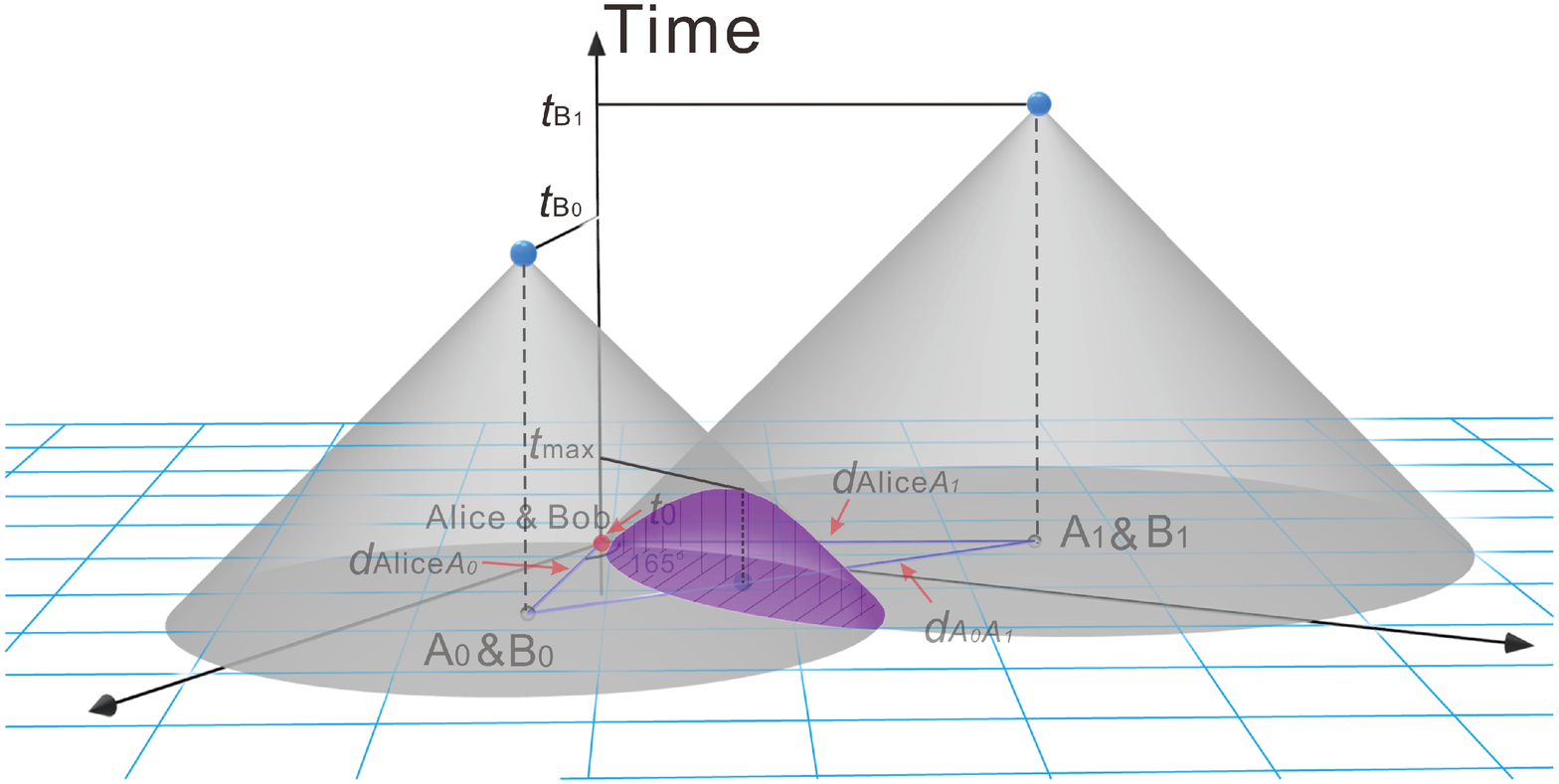}
 \end{center}
 \caption{Space-time diagram of the experiment. Alice can only have made her commitment in the intersection (shaded part) of $B_0$ and $B_1$'s past light cones, given that they have received the signals from $A_0$ and $A_1$. This area is in magenta in the figure. The latest time instant where Alice could have committed is given by $t_{\rm max}$; its projection in space lies on the line connecting $A_0$ with $A_1$.}
 \label{spacetime}
\end{figure}

In the verification step of the protocol it is also important to determine the latest time instant in which Alice could have made her commitment, given that Bob accepted the revealed bit. We denote this quantity as $t_{\rm commit}$. From the geographical distribution of the different parties involved in the protocol,
it is straightforward to obtain an upper bound for this quantity. This is illustrated in Fig.~\ref{spacetime}. Here, $d_{ij}$ denotes the distance between parties $i$ and $j$ in the protocol, $t_0$ represents the time instant where Bob sends Alice his first signal, and $t_{B_0}$ ($t_{B_1}$) is the time instant where agent $B_0$ ($B_1$) receives the last signal from agent $A_0$ ($A_1$). These parameters are directly observed in the protocol.
Furthermore, suppose, for the moment, that $d_{A_0B_0},d_{A_1B_1}, d_{\rm AliceBob}, \ll d_{{\rm Alice}A_0},d_{{\rm Alice}A_1},d_{A_0A_1}$ and $N_{\rm tol}$ is large (to guarantee
a small cheating probability) and thus the total number of signals $N$
sent by Bob is also large. In this scenario, it can be shown that
\begin{equation}\label{commit}
t_{\rm commit}\leq{}t_{\rm max}\equiv{}\frac{1}{2}\left(t_{B_0}+t_{B_1}-\frac{d_{A_0A_1}}{c}\right)-t_0,
\end{equation}
where $c$ denotes the speed of light in vacuum. A proof of a more general version of this statement can be found in Appendix~C.

The protocol can also guarantee that the commitment is not performed in certain space points, {\it e.g.}, in the locations of agents $A_0$ and $A_1$.
For this, Bob may verify the conditions
$d_{{\rm Alice}A_i}+d_{A_0A_1}>c(t_{B_{i\oplus1}}-t_0)$, with $i=0,1$,
which assures the latter.

%%%%%%%%%%%%%%%%%%%%%%%%%%%%%%%%%%%%%%%%%%%%%%%%%%%%%%%%%%%%%%%%%%%
% Experimental part
%%%%%%%%%%%%%%%%%%%%%%%%%%%%%%%%%%%%%%%%%%%%%%%%%%%%%%%%%%%%%%%%%%%

We performed a field test of the protocol among the three geographically separated laboratories shown in Fig.~\ref{diagram}a.
One important detail to consider in the experiment is that a higher transmission speed reduces the earliest time where Alice may reveal her committed bit, and thus it can
also reduce the value of $t_{B_0}$ and $t_{B_1}$. According to Eq.~(\ref{commit}), this also decreases $t_{\rm commit}$. The optical communication speed in a free-space channel is $1.5$ times higher than in a fiber channel. Therefore, in our experiment, we choose a free-space channel for the communication between Alice and her agents. The distance between Alice/Bob's lab and $A_0/B_0$'s lab is about 9.3 km, and the distance between Alice/Bob's lab and $A_1/B_1$'s lab is about $12.3$ km. The angle of $A_0$-Alice-$A_1$ is around $165$ degrees.

When an experimental run starts, triggered by a GPS signal, Bob randomly prepares phase-randomised weak coherent pulses in four different polarisation states and sends them to Alice. As shown in Fig.~\ref{diagram}b, the random pulsed optical signals are emitted from four diodes at a repetition frequency of $50$ MHz. These diodes are controlled by random numbers generated off line by quantum random number generators (QRNGs). The central wavelength of all laser diodes is $850$ nm, and the average photon number is adjusted to $0.183 \pm 10\%$ per pulse. In order to send more signals within a certain time interval, we utilise two parallel BB84 systems in the experiment. In each run, Bob sends two sequences of $2838$ pulses within $56.76$ $\mu$s. The delay between the time when Bob sends his first signal and the triggered GPS signal is measured as $1.53$ $\mu$s, which is taken as the initial time, $t_0$.

Alice uses a HWP to choose the measurement basis and two SPDs to implement the measurement. When she selects the rectilinear basis, bit $0$ is committed, whereas when she chooses the diagonal basis, bit $1$ is committed. The detection efficiency, dark count rate and dead time of each SPD are, respectively, $50\%$, $100$ cps and $30$ ns. The total detection efficiency of the measurement setup is around $45\%$, including a transmission and collection efficiency of $90\%$ together with the SPD's detection efficiency. An FPGA board is used to record and process the detection information. When one detector clicks, the FPGA board records which detector has clicked and the time instant when this happened. When both detectors click, the FPGA board records the information of one of them randomly chosen. If there are two detection events in a $60$ ns time interval, the later detection event's information is dropped. This last procedure is implemented to keep a dishonest Bob from attacking Alice's detection device. Meanwhile, the FPGA board also sends all the detection timing information to Bob through 1 GHz optical communication. This step is needed to keep a dishonest Alice from cheating~\cite{Kent12}.

Next, Alice encrypts all her data using the OTP and sends it to her agents $A_0$ and $A_1$ via 1 GHz optical communication. The secret keys $K_{A_0}$ and $K_{A_1}$ for OTP have been generated using a QRNG and shared between Alice and her agents off line. In order to communicate through a long distance free-space channel, Alice uses erbium doped fiber amplifiers (EDFAs) to amplify the optical signals to an average power of $200$ mW for the Alice-$A_0$ channel, and $1$ W for the Alice-$A_1$ channel. Kepler telescopes with aperture of $80$ mm and $127$ mm are used to send the amplified signals to agents $A_0$ and $A_1$, respectively. Cassegrain telescopes with aperture of $150$ mm are used by both agents to receive the signals. In order to achieve a stable and highly efficient free-space optical channel, we employ the acquiring, pointing and tracking (APT) technique in both the transmitter and the receiver. The optical signals are then collected into a multi-mode fiber with a diameter of $62.5$ $\mu$m. At the output of the fiber, we observe an average power of more than $100$ $\mu$W, which is high enough for a classical optical detector.

$A_0$ and $A_1$ decrypt the optical signals using their own respective secret keys as illustrated in Fig.~\ref{diagram}c. After receiving all the data sent by Alice, they forward the decrypted information to agents $B_0$ and $B_1$ via 1 GHz optical communication, respectively, to unveil the committed bit value. Then, $B_0$ and $B_1$ compare the information received together with Bob. If the data sent by $A_0$ and $A_1$ is not equal, Bob rejects the commitment. Otherwise, he calculates the parameters 
$n_{\rm rect}$, $n_{\rm diag}$, $n_{\rm e, rect}$ and $n_{\rm e, diag}$ following the procedure described in Appendix~B. Only when these parameters satisfy the conditions described in the verification step of the protocol, Bob accepts the commitment. Meanwhile, with the help of their own GPSs, $B_0$ and $B_1$ record the arrival time of the signals sent by $A_0$ and $A_1$. Based on this timing information, Bob and his agents determine $t_{\rm commit}$ according to Eq.~(\ref{commit}). All the communications
between Alice's and Bob's agents use a bandwidth of 1 GHz.

%%%%%%%%%%%%%%%%%%%%%%%%%%%%%%%%%%%%%%%%%%%%%%%%%%%%%%%%%%%%%%%%%%%
%Experimental results
%%%%%%%%%%%%%%%%%%%%%%%%%%%%%%%%%%%%%%%%%%%%%%%%%%%%%%%%%%%%%%%%%%%

We performed the experiment $8$ times, in half of which Alice commits to the bit value $0$ and in the other half she commits to
$1$. The results are shown in Tab.~\ref{Tab:Rect2838}. In each run, Alice detects around $400$ pulses. The total bit error rate is around $1\%$ when the commitment basis coincides with the preparation basis. This error is mainly due to the optical baseline error and the detector's dark counts.

The time interval between commit and unveil is about $30$ $\mu$s for all the trails. As unveiling time, which we denote as $t_{\rm unveil}$, we consider the instant where $A_0$ sends the first signal to $B_0$, since in our experiment this always happens before $A_1$ sends a signal to $B_1$. From these results, Bob can also conclude that the commitment was not done in the locations of $A_0$ or $A_1$. 

Both quantum mechanics and relativity have changed our understanding of the universe. Our experiment shows for the first time that when we combine them
we can solve a fundamental problem with many practical applications, and for which there is no solution using only one of them on their own. Our work demonstrates that quantum relativistic communication is experimentally feasible, and opens a promising new field for research with technological applications.

\textbf{\subsection*{Acknowledgments}}
We acknowledge insightful discussions with Xiang-Bin Wang, Chang Liu, Jing Lin, Yang Li, Ya-Li Mao, and Vicente Martin. This work has been supported by the National Fundamental Research Program (under Grant No. 2013CB336800, 2011CB921300 and 2011CBA00300), the NNSF of China, the CAS. AC was supported by the Project No.\ FIS2011-29400 (MINECO, Spain). MC acknowledges support from the European Regional Development Fund (ERDF), the Galician Regional Government (projects CN2012/279 and CN 2012/260, Consolidation of Research Units: AtlantTIC).
%\textbf{\subsection*{Author Contributions}}
%All authors contributed extensively to the work presented in this paper.
%\textbf{\subsection*{Author Information}}
%The authors declare no competing financial interests.

%%%%%%%%%%%%%%%%%%%%%%%%%%%%%%%%%%%%%%%%%%%%%%%%%%%%%%%%%%%%%%%%%%%

\bibliographystyle{apsrev4-1}
\bibliography{Bibli_v7}
\clearpage

%%%%%%%%%%%%%%%%%%%%%%%%%%%%%%%%%%%%%%%%%%%%%%%%%%%%%%%%%%%%%%%%%%%
% Fig. 1
%%%%%%%%%%%%%%%%%%%%%%%%%%%%%%%%%%%%%%%%%%%%%%%%%%%%%%%%%%%%%%%%%%%

\begin{figure}[htp]
  \captionsetup{labelformat=empty}
  \centering
  \includegraphics[scale=0.82]{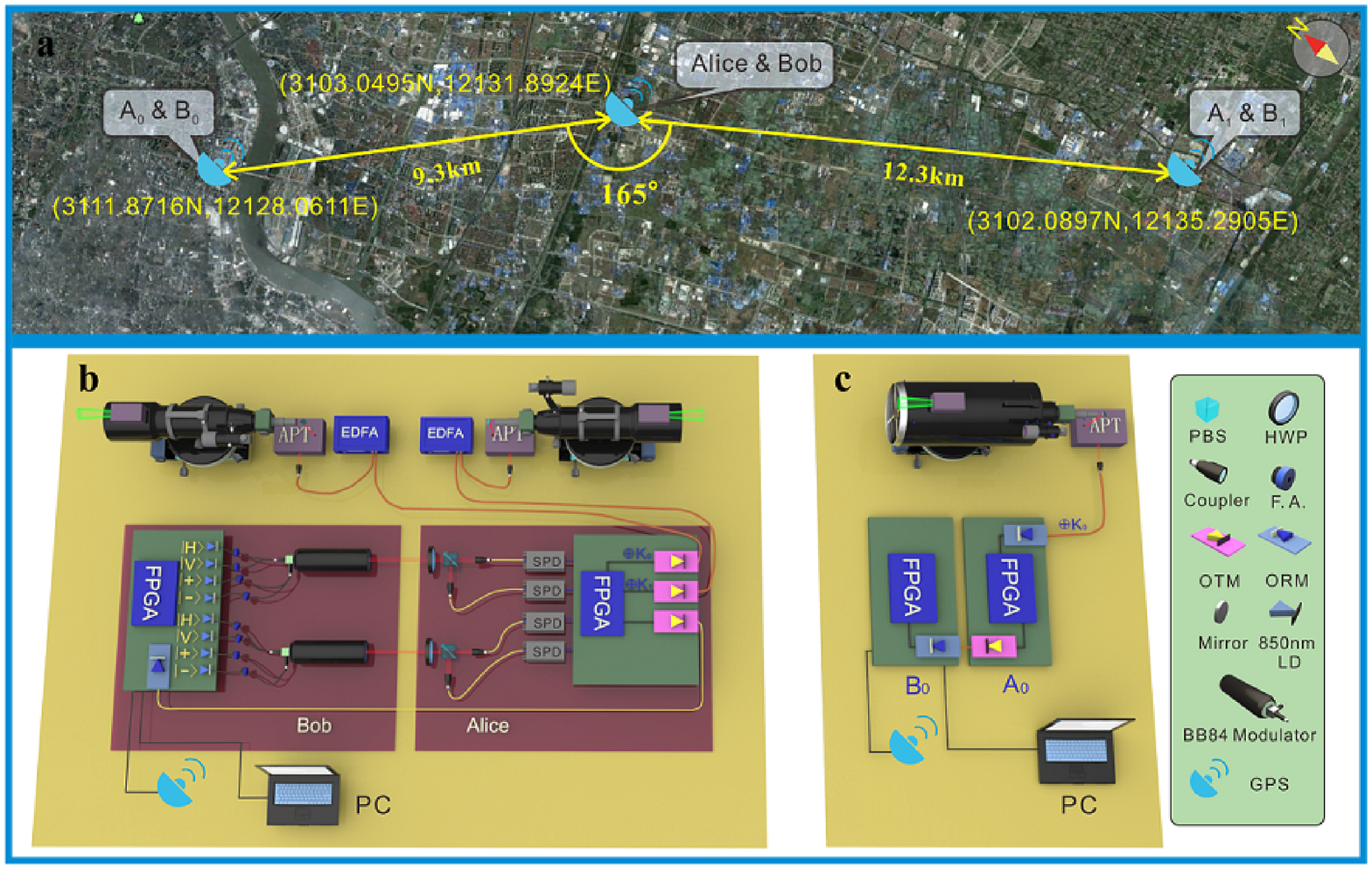}
  \caption{}
\end{figure}
\clearpage
\begin{figure}[htp]
  \captionsetup{labelformat=adja-page}
  \ContinuedFloat
 \caption{
 (a) Diagram of the geographical distribution of the parties. The distances between the three main locations are indicated in the figure. The distance between Alice and Bob, and between $A_0$ and $B_0$ (and $A_1$ and $B_1$) is less than one meter. Alice communicates with her agents using sending and receiving telescopes.
 Each site is equipped with a global position system (GPS) for synchronisation.
 (b) Diagram of Bob's and Alice's setup. Triggered by a GPS signal, Bob attenuates and encodes laser pulses with a BB84~\cite{BB84} module, which is composed of two Wollason polarisation prisms and a beam splitter, and sends them to Alice. Alice's commitment setup consists of a half wave plate (HWP), a polarisation beam splitter (PBS) and silicon avalanched photo-diode single photon detectors (SPDs). A field programmable gate array (FPGA) board is used to record the detected signals and communicate with Bob and Alice's agents. Alice encodes, amplifies and sends the measurement results to her agents through telescopes.
 The four BB84 polarisation states are denoted in the figure as $\ket{{\rm H}}$, $\ket{{\rm V}}$, $\ket{{\rm +}}$ and $\ket{{\rm -}}$.
 F.A.: fixed attenuator. OTM: optical transmission module. ORM: optical receiving module. EDFA: erbium doped fiber amplifies. APT: acquiring, pointing and tracking.
 (c) Diagram of $B_0$'s and $A_0$'s setup.
 The one of $B_1$ and $A_1$ is identical.
 $A_0$'s FPGA board receives and decrypts the detection information, stores it until the last classical signal is received, and then sends the measurement results to $B_0$. Bob's agents record the timing of these signals and send the results to Bob.}
  \label{diagram}
\end{figure}
\clearpage

%%%%%%%%%%%%%%%%%%%%%%%%%%%%%%%%%%%%%%%%%%%%%%%%%%%%%%%%%%%%%%%%%%%
% Fig. 2
%%%%%%%%%%%%%%%%%%%%%%%%%%%%%%%%%%%%%%%%%%%%%%%%%%%%%%%%%%%%%%%%%%%

%%%%%%%%%%%%%%%%%%%%%%%%%%%%%%%%%%%%%%%%%%%%%%%%%%%%%%%%%%%%%%%%%%%
%Tables
%%%%%%%%%%%%%%%%%%%%%%%%%%%%%%%%%%%%%%%%%%%%%%%%%%%%%%%%%%%%%%%%%%%

\clearpage
\begin{table}
\centering
\begin{tabular}{c|cccccccc}
\hline
\hline
 & Exp $1$ & Exp $2$ & Exp $3$ & Exp $4$ & Exp $5$ & Exp $6$ & Exp $7$ & Exp $8$  \\
\hline
Bit committed   & $0$ & $1$ & $0$ & $0$ & $1$ & $0$ & $1$ & $1$ \\
Bit deduced     & $0$ & $1$ & $0$ & $0$ & $1$ & $0$ & $1$ & $1$ \\
$n_{\rm e, i}$     & $0$ & $1$ & $1$ & $0$ & $0$ & $0$ & $0$ & $0$ \\
$n_{\rm rect}$   & $112$ & $119$ & $126$ & $118$ & $115$ & $109$ & $109$ & $119$ \\
$n_{\rm diag}$   & $116$ & $120$ & $107$ & $128$ & $115$ & $122$ & $141$ & $150$\\
$n_{\rm e, i}/N_{\rm tol}$  & $0\%$ & $0.93\%$ & $0.93\%$ & $0\%$ & $0\%$ & $0\%$   & $0\%$ & $0\%$ \\
$t_{B_0}$ ($\mu$s) & $92.85$ & $93.02$ & $92.99$ & $92.98$ & $93.18$ & $92.97$ & $93.12$ & $93.24$ \\
$t_{B_1}$ ($\mu$s) & $102.74$ & $102.85$ & $102.92$ & $102.93$ & $103.22$ & $102.84$ & $103.08$ & $103.10$ \\
 $t_{\rm commit}$ ($\mu$s)   & $60.54$ & $60.68$ & $60.70$ & $60.70$ & $60.94$ & $60.65$ & $60.84$ & $60.91$ \\
$t_{\rm unveil}$ ($\mu$s) & $90.57$ & $90.62$ & $90.63$ & $90.58$ & $90.62$ & $90.53$ & $90.68$ & $90.68$ \\
\hline
\hline
\end{tabular}
\caption{Experimental results when Bob sends Alice two parallel sequences of $2838$ pulses. The parameter 
$n_{\rm e, i}$ refers to $n_{\rm e, rect}$ ($n_{\rm e, diag}$) when Alice commits to a bit value $0$ ($1$).
In these experiments we fix the tolerated values $N_{\rm tol}=107$ and $E_{\rm tol}=1.5\%$, and obtain
a cheating probability below
$5.68\times10^{-2}$.}
\label{Tab:Rect2838}
\end{table}

%%%%%%%%%%%%%%%%%%%%%%%%%%%%%%%%%%%%%%%%%%%%%%%%%%%%%%%%%%%%%%%%%%%

\clearpage

%%%%%%%%%%%%%%%%%%%%%%%%%%%%%%%%%%%%%%%%%%%%%%%%%%%%%%%%%%%%%%%%%%%

\appendix

\section{Security analysis}

In this Appendix we analyse the security of the bit commitment protocol
implemented. For this,
we use the security proof technique introduced in~\cite{KTHW12}. However, while~\cite{KTHW12} considers an
error-free case and assumes that Bob sends Alice single-photon pulses, here
we analyse the practical situation where the signals prepared by Bob are phase-randomised weak coherent pulses
and the error rate of the single-photon contributions is below a certain prefixed value $E_{\rm tol}$ (see the definition of
the protocol in the paper).

We begin by introducing some technical definitions.
A bit commitment protocol is
$\epsilon_{\rm c}$-concealing if Bob cannot learn any information about the committed bit $b$
before Alice unveils it, except with a minuscule probability $\epsilon_{\rm c}$. And,
it is
$\epsilon_{\rm b}$-binding if Bob has a guarantee that~\cite{KTHW12,dumais}
$p_0+p_1\leq{}1+\epsilon_{\rm b}$, where $p_0$ ($p_1$) represents the probability that Bob
accepts Alice's commitment to be $0$ ($1$).
Note that the binding condition in quantum bit commitment protocols
is slightly different from that used in classical schemes, which typically requires that
either $p_0$ or $p_1$ is very small after the
commit phase. See~\cite{KTHW12,dumais} for a detailed discussion related to this issue.
We say that a commitment is $\epsilon$-secure, with $\epsilon_{\rm c}+\epsilon_{\rm b}\leq\epsilon$, if it is
$\epsilon_{\rm c}$-concealing and $\epsilon_{\rm b}$-binding.

In the next two sections we demonstrate that the bit commitment protocol implemented~\cite{Kent12} is
perfectly concealing ({\it i.e.}, $\epsilon_{\rm c}=0$) and $\epsilon_{\rm b}$-binding,
with $\epsilon_{\rm b}$ given by Eq.~(\ref{main}). We begin by proving
its security against a dishonest Alice.

\subsection{Security against a dishonest Alice:}
The main technical result of this section is Claim 1 below. It states that the bit
commitment protocol considered in the paper is $\epsilon_{\rm b}$-binding, with
$\epsilon_{\rm b}$ approximating
zero when $N_{\rm tol}$ increases, given that
the tolerated value $E_{\rm tol}$ is not too large. This result applies to the general global command model
introduced in~\cite{KTHW12,kentb}, where it is assumed that Alice's agents $A_0$ and $A_1$ may receive a global command to decide
which bit value unveil. For instance, $A_0$ and $A_1$ could decide to
reveal either $0$ or $1$ depending on some global news simultaneously available to both of them.

\noindent{\bf Claim 1:} {\it The bit commitment protocol described in the paper
is $\epsilon_{\rm b}$-binding, with}
\begin{eqnarray}\label{main}
\epsilon_{\rm b}&\leq& \inf_{\delta\in(E_{\rm tol},1/2)}
\Bigg\{\left[1-\exp\left(\frac{(\delta{}N_{\rm tol} -\lfloor{}E_{\rm tol}N_{\rm tol}\rfloor)^2}{1-N_{\rm tol}}\right)\right]
2^{1-[1-h(\delta)]N_{\rm tol}} \\
&&\quad
+2\exp\left(\frac{(\delta{}N_{\rm tol} -\lfloor{}E_{\rm tol}N_{\rm tol}\rfloor)^2}{1-N_{\rm tol}}\right)
\Bigg\}
\left[1+\sum_{k=1}^{\lfloor E_{\rm tol}N_{\rm tol}\rfloor}(2^k-1){N_{\rm tol} \choose k}\right] +\varepsilon_{\rm rect}+\varepsilon_{\rm diag}, \nonumber
\end{eqnarray}
{\it  where the function $h(x)=-x\log_2{(x)}-(1-x)\log_2{(1-x)}$ is the binary Shannon entropy function,
$E_{\rm tol}$ denotes the tolerated error rate of the protocol, $N_{\rm tol}$ represents the minimum number of single photons
prepared in the rectilinear basis (and also in the diagonal basis)
that Alice needs to detect, and $\varepsilon_{\rm rect}$ and $\varepsilon_{\rm diag}$
are the probabilities that the estimation of the terms $n_{\rm rect}$ and $n_{\rm diag}$ is incorrect.}

\begin{proof}
The first fact to notice is that all multi-photon pulses sent by Bob
are insecure. This is so because a dishonest Alice may perform a quantum non-demolition
measurement of the total number of photons contained in each signal. Whenever she observes a
multi-photon state, she can measure one photon in the rectilinear basis and another photon
in the diagonal basis. Then she sends both results to her agents $A_0$ and $A_1$. With this information,
and assuming the global
command model,
$A_0$ and $A_1$ can always make $p_0=p_1=1$.
From now on, therefore, we will consider only the
single-photon states sent by Bob and detected by Alice. These are the only
contributions that can make the security parameter $\epsilon_{\rm b}$ close to zero.

To prove the security of the single-photon pulses sent by Bob we consider a virtual qubit idea.
Instead of preparing a single-photon BB84 state, Bob prepares its
purification. That is, one can think of Bob actually having a qubit on his side.
Then, he generates a signal by first preparing an entangled state of the combined
system of his virtual qubit and the qubit that he is sending Alice in say a singlet state.
He subsequently measures his virtual qubit, thus preparing a BB84 state.
This virtual scheme is completely equivalent to the original one in terms of its security. More precisely,
we will consider that
Bob prepares $n_{\rm rect}+n_{\rm diag}$ singlet states and sends one
qubit from each of these states to Alice, while
he keeps the other qubit.
Now, in principle, Bob may
keep his $n_{\rm rect}+n_{\rm diag}$ virtual qubits in a quantum memory and
delay his measurement on them.
Only after Alice's agents
$A_0$ and $A_1$ have given all their results to agents $B_0$ and $B_1$,
Bob selects at random $n_{\rm rect}$ virtual qubits and measures them in the rectilinear basis.
Likewise, he measures the remaining $n_{\rm diag}$ qubits in the
diagonal basis.

Now, we need to introduce some further notations~\cite{KTHW12}.
Let $\rho_{BA_0A_1}$ denote the quantum state shared by Bob and the agents $A_0$ and $A_1$
before the commitment is revealed. Also, let $\Phi_{A_0}^b$ ($\Phi_{A_1}^b$) represent
the map applied by agent $A_0$ ($A_1$) with the intention to open the bit value $b$.
Importantly,
the map
$\Phi_{A_0}^b$ ($\Phi_{A_1}^b$) is restricted to act only on the subsystem hold by $A_0$ ($A_1$).
These two maps produce respectively the
output bit strings $S_{A_0}$ and $S_{A_1}$, which are
given to agents $B_0$ and $B_1$.
As described above, only after $B_0$ and $B_1$ have received, respectively, $S_{A_0}$ and $S_{A_1}$,
Bob decides which virtual qubits he measures in
the rectilinear basis and which ones are measured in the diagonal basis. In so doing,
we can naturally split
the bit string $S_{A_0}$ into
two substrings $S_{A_0}=\{S_{A_0,{\rm rect}},S_{A_0,{\rm diag}}\}$, where
$S_{A_0,{\rm rect}}$ ($S_{A_0,{\rm diag}}$) contains those bits
of $S_{A_0}$ associated with events where Bob measures the corresponding virtual qubit in the rectilinear (diagonal) basis.
Likewise, Bob does the same with the bit string
$S_{A_1}$.
Also, we split Bob's system into $B_{\rm rect}$ and $B_{\rm diag}$.
The first (second)
subsystem represents those virtual qubits that Bob measures in the rectilinear (diagonal) basis.
That is, we have $|B_{\rm rect}|=n_{\rm rect}$ and $|B_{\rm diag}|=n_{\rm diag}$.
Moreover, let the quantum operation $\Lambda_{G,\beta}$, with $G\in\{B_{\rm rect},B_{\rm diag}\}$ and
$\beta\in\{\rm rect, diag\}$,
correspond to measuring all qubits from subsystem $G$ using the basis $\beta$.
That is, $\Lambda_{B_{\rm rect},{\rm rect}}$ and $\Lambda_{B_{\rm diag},{\rm diag}}$ denote
the measurements
implemented by Bob in the virtual protocol. The quantum operation
$\Lambda_{B_{\rm diag},{\rm rect}}$ is not performed in the protocol.
However,
we will use it for the purposes of
the security proof. The result of applying
$\Lambda_{B_{\rm rect},{\rm rect}}$ ($\Lambda_{B_{\rm diag},{\rm diag}}$) to Bob's subsystem
$B_{\rm rect}$ ($B_{\rm diag}$) is a bit string that we shall denote as $S_{B_{\rm rect}}$ ($S_{B_{\rm diag}}$).
Finally, let $\Pi_{A_0}^b$ ($\Pi_{A_1}^b$) be the operation that Bob uses to check if the results declared
by agent $A_0$ ($A_1$) are consistent with committing to a bit value $b$.

Using precisely the same arguments of~\cite{KTHW12}, it is easy to show that
the security parameter $\epsilon_{\rm b}$ is upper bounded by the probability that $A_0$ tries to unveil the bit value $0$, $A_1$ tries to
unveil the bit value $1$, and both results are accepted by Bob given that he makes a separately
decision for each of these two agents. Next, we calculate an upper bound for this probability.
For this,
let $\sigma=\rho_{S_{B_{\rm rect}}S_{B_{\rm diag}}S_{A_0,{\rm rect}}S_{A_0,{\rm diag}}S_{A_1,{\rm rect}}S_{A_1,{\rm diag}}}$
be the classical state after Bob, $A_0$ and $A_1$ have made all their measurements (with $A_i$ trying to unveil the bit value $i$,
with $i=0,1$), {\it i.e.},
\begin{equation}
\sigma=
\left(\Lambda_{B_{\rm rect},{\rm rect}}\otimes
\Lambda_{B_{\rm diag},{\rm diag}}\otimes
\Phi_{A_0}^0\otimes\Phi_{A_1}^1\right)\rho_{B_{\rm rect}B_{\rm diag}A_0A_1},
\end{equation}
where we already used the fact that $B=B_{\rm rect}B_{\rm diag}$.
With this notation, we have that $\epsilon_{\rm b}$ can be expressed as
\begin{equation}\label{ep}
\epsilon_{\rm b}\leq{\rm tr}\left(\Pi_{A_0}^0\Pi_{A_1}^1\sigma\right).
\end{equation}
Now, in order to evaluate Eq.~(\ref{ep}), we introduce two further quantities.
In particular, let
$p_{A_0}$ be the probability that $A_0$ passes the test, and let
$\rho_{B_{\rm diag}S_{A_0,{\rm diag}}A_1}^{\rm pass}$ be the state
conditioned on passing. That is,
\begin{eqnarray}
p_{A_0}&=&{\rm tr}\left[\Pi_{A_0}^0\left(\Lambda_{B_{\rm rect},{\rm rect}}\otimes\Phi_{A_0}^0\right)
\rho_{B_{\rm rect}B_{\rm diag}A_0A_1}\right], \\
\rho_{B_{\rm diag}S_{A_0,{\rm diag}}A_1}^{\rm pass}&=&
\frac{1}{p_{A_0}}{\rm tr}_{S_{B_{\rm rect}}S_{A_0,{\rm rect}}}
\left[\Pi_{A_0}^0(\Lambda_{B_{\rm rect},{\rm rect}}\otimes\Phi_{A_0}^0)\rho_{B_{\rm rect}B_{\rm diag}A_0A_1}\right]. \nonumber
\end{eqnarray}
This means that Eq.~(\ref{ep}) can be equivalently written as
\begin{equation}\label{ep1}
\frac{\epsilon_{\rm b}}{p_{A_0}}\leq{\rm tr}
\left[\Pi_{A_1}^1(\Lambda_{B_{\rm diag},{\rm diag}}\otimes\Phi_{A_1}^1)\rho_{B_{\rm diag}S_{A_0,{\rm diag}}A_1}^{\rm pass}\right].
\end{equation}

The term on the r.h.s. of Eq.~(\ref{ep1}) represents the probability that $A_1$ passes the test
conditioned on Bob accepting the result declared by $A_0$. In order for $A_1$ to pass
the test, we need that
$n_{\rm e,{\rm diag}}/N_{\rm tol}\leq{}E_{\rm tol}$ (see the definition of the protocol in the paper). This
condition is equivalent to require that the
Hamming distance between $S_{B_{\rm diag}}$ and $S_{A_1,{\rm diag}}$ is
less or equal than $n_{\rm e,{\rm diag}}\leq\lfloor E_{\rm tol}N_{\rm tol}\rfloor$.
That is, to pass the test $A_1$ needs to
correctly guess at least $n_{\rm diag}-\lfloor E_{\rm tol}N_{\rm tol}\rfloor$ bits from
the bit string
$S_{B_{\rm diag}}$.
Then, using a result from~\cite{guess} we can obtain a simple upper for the probability that $A_1$ passes the test,
\begin{eqnarray}\label{vir}
{\rm tr}
\left[\Pi_{A_1}^1(\Lambda_{B_{\rm diag},{\rm diag}}\otimes\Phi_{A_1}^1)\rho_{B_{\rm diag}S_{A_0,{\rm diag}}A_1}^{\rm pass}\right]
&\leq&{}\left[1+\sum_{k=1}^{\lfloor E_{\rm tol}N_{\rm tol}\rfloor}(2^k-1){n_{\rm diag} \choose k}\right] \nonumber \\
&\times&2^{-H_{\rm min}(S_{B_{\rm diag}}|A_1)},
\end{eqnarray}
where $H_{\rm min}(S_{B_{\rm diag}}|A_1)$ denotes the conditional min-entropy
evaluated on
the state
$\rho_{S_{B_{\rm diag}}A_1}={\rm tr}_{S_{A_0,{\rm diag}}}(\Lambda_{B_{\rm diag},{\rm diag}}
\rho_{B_{\rm diag}S_{A_0,{\rm diag}}A_1}^{\rm pass})$.
To prove Eq.~(\ref{vir}) note that
\begin{equation}\label{uno}
{\rm tr}
\left[\Pi_{A_1}^1(\Lambda_{B_{\rm diag},{\rm diag}}\otimes\Phi_{A_1}^1)\rho_{B_{\rm diag}S_{A_0,{\rm diag}}A_1}^{\rm pass}\right]
=\sum_{k=0}^{\lfloor E_{\rm tol}N_{\rm tol}\rfloor}{\rm pr}\left[d_{\rm H}(S_{B_{\rm diag}},S_{A_1,{\rm diag}})=k\right],
\end{equation}
where $d_{\rm H}(x,y)$ denotes the Hamming distance between the bit strings $x$ and $y$.
From~\cite{guess} we have that 
${\rm pr}\left[d_{\rm H}(S_{B_{\rm diag}},S_{A_1,{\rm diag}})=0\right]\leq2^{-H_{\rm min}(S_{B_{\rm diag}}|A_1)}$. 
Similarly, let ${\hat S}_{B_{\rm diag},n_{\rm diag}-k}$ denote a substring of $S_{B_{\rm diag}}$ of size $n_{\rm diag}-k$, 
with $k>0$. The probability 
that $A_1$ guesses correctly ${\hat S}_{B_{\rm diag},n_{\rm diag}-k}$ and fails in the remaining $k$ bits of $S_{B_{\rm diag}}$ is 
upper bounded by
\begin{equation}
2^{-H_{\rm min}({\hat S}_{B_{\rm diag},n_{\rm diag}-k}|A_1)}-2^{-H_{\rm min}(S_{B_{\rm diag}}|A_1)}
\leq2^{-H_{\rm min}(S_{B_{\rm diag}}|A_1)}\left(2^k-1\right), 
\end{equation}
where in the inequality we have used the fact that 
$H_{\rm min}({\hat S}_{B_{\rm diag},n_{\rm diag}-k}|A_1)\geq H_{\rm min}(S_{B_{\rm diag}}|A_1)-k$. Then, 
if we take into account all the possible substrings ${\hat S}_{B_{\rm diag},n_{\rm diag}-k}$ contained in 
$S_{B_{\rm diag}}$, we have that
\begin{equation}\label{dos}
{\rm pr}\left[d_{\rm H}(S_{B_{\rm diag}},S_{A_1,{\rm diag}})=k\right]
\leq
2^{-H_{\rm min}(S_{B_{\rm diag}}|A_1)}\left(2^k-1\right){n_{\rm diag} \choose k},
\end{equation}
for $k>0$.
Combining Eqs.~(\ref{uno})-(\ref{dos}) we obtain Eq.~(\ref{vir}).

Now, we employ the uncertainty relation introduced in \cite{tom}. It states that
\begin{equation}\label{un}
H_{\rm max}({\tilde S}_{B_{\rm diag}}|S_{A_0,{\rm diag}})+H_{\rm min}(S_{B_{\rm diag}}|A_1)\geq{}n_{\rm diag},
\end{equation}
where $H_{\rm max}({\tilde S}_{B_{\rm diag}}|S_{A_0,{\rm diag}})$ 
represents the max-entropy evaluated
on the state
\begin{equation}
\rho_{{\tilde S}_{B_{\rm diag}}S_{A_0,{\rm diag}}}=
{\rm tr}_{A_1}\left(\Lambda_{B_{\rm diag},{\rm rect}}
\rho_{B_{\rm diag}S_{A_0,{\rm diag}}A_1}^{\rm pass}\right).
\end{equation}
As already mentioned previously, note that
the operation $\Lambda_{B_{\rm diag},{\rm rect}}$ is not performed
in the protocol. However, we can estimate its result.
Combining this result with
Eqs.~(\ref{ep1})-(\ref{vir}) we find that
\begin{equation}\label{gim}
\frac{\epsilon_{\rm b}}{p_{A_0}}\leq 
\left[1+\sum_{k=1}^{\lfloor E_{\rm tol}N_{\rm tol}\rfloor}(2^k-1){n_{\rm diag} \choose k}\right]
2^{H_{\rm max}\left({\tilde S}_{B_{\rm diag}}|S_{A_0,{\rm diag}}\right)-n_{\rm diag}}.
\end{equation}

The next step is to evaluate the quantity $H_{\rm max}({\tilde S}_{B_{\rm diag}}|S_{A_0,{\rm diag}})$.
For this, let $n_{\rm e, rect}$ denote the total number of errors detected in the declaration of $A_0$ (in the
rectilinear basis) conditioned
on passing.
That is, we have that $n_{\rm e, rect}\leq{}\lfloor E_{\rm tol}N_{\rm tol}\rfloor$.
Then, using Serfling inequality~\cite{serf}
for random sampling without replacement we find that
\begin{eqnarray}\label{last}
&&{\rm pr}\left[d_{\rm H}\left({\tilde S}_{B_{\rm diag}},S_{A_0,{\rm diag}}\right)\geq \delta n_{\rm diag}|
d_{\rm H}\left(S_{B_{\rm rect}},S_{A_0,{\rm rect}}\right)=n_{\rm e, rect}\right] \nonumber \\
&&\leq \exp\left\{\frac{-[\sqrt{2}(\delta n_{\rm rect}-n_{\rm e, rect})n_{\rm diag}]^2}
{n_{\rm rect}(n_{\rm diag}-1)(n_{\rm rect}+n_{\rm diag})}\right\}\equiv\frac{\gamma}{p_{A_0}},
\end{eqnarray}
with $\delta\in(E_{\rm tol},1/2)$.
That is, Eq.~(\ref{last}) represents an upper bound for the probability of finding more than
$\delta n_{\rm diag}$ errors between the bit strings ${\tilde S}_{B_{\rm diag}}$ and $S_{A_0,{\rm diag}}$ given
that we observed $n_{\rm e, rect}$ errors between the bit strings $S_{B_{\rm rect}}$ and $S_{A_0,{\rm rect}}$.

Now, we define the binary event $\Gamma$ as
\begin{equation}
\Gamma = \left\{ \begin{array}{ll}
 0 & \textrm{if $d_{\rm H}({\tilde S}_{B_{\rm diag}},S_{A_0,{\rm diag}})< \delta n_{\rm diag}$},\\
1 & \textrm{if $d_{\rm H}({\tilde S}_{B_{\rm diag}},S_{A_0,{\rm diag}})\geq \delta n_{\rm diag}$},
  \end{array} \right.
\end{equation}
and we use the same techniques employed in~\cite{KTHW12}. In particular, it can be shown that
\begin{equation}
2^{H_{\rm max}({\tilde S}_{B_{\rm diag}}|S_{A_0,{\rm diag}},\Gamma)}\leq
\left(1-\frac{\gamma}{p_{A_0}}\right)
2^{n_{\rm diag}h(\delta)}+2^{n_{\rm diag}}\frac{\gamma}{p_{A_0}},
\end{equation}
where $h(\cdot)$ denotes again the binary Shannon entropy function.
Now, from \cite{Tomamichel2012thesis} we have that
\begin{equation}
H_{\rm max}({\tilde S}_{B_{\rm diag}}|S_{A_0,{\rm diag}})
\leq H_{\rm max}({\tilde S}_{B_{\rm diag}}|S_{A_0,{\rm diag}},\Gamma)+1.
\end{equation}
Combining these results with Eq.~(\ref{gim}), we obtain that
\begin{eqnarray}
\epsilon_{\rm b}\leq 
&&\inf_{\delta\in(E_{\rm tol},1/2)}
\left[\left(1-\frac{\gamma}{p_{A_0}}\right)
2^{1-[1-h(\delta)]n_{\rm diag}}+2\frac{\gamma}{p_{A_0}}\right] \nonumber \\
&&\times\left[1+\sum_{k=1}^{\lfloor E_{\rm tol}N_{\rm tol}\rfloor}(2^k-1){n_{\rm diag} \choose k}\right],
\end{eqnarray}
where the parameter $\gamma/p_{A_0}$ is given by Eq.~(\ref{last}). 
Finally, if we take into account that $n_{\rm e, rect}\leq{}\lfloor E_{\rm tol}N_{\rm tol}\rfloor$ and,
moreover, that 
when $A_0$ and $A_1$ pass the test then
$n_{\rm rect}, n_{\rm diag}\geq{}N_{\rm tol}$, we obtain
\begin{eqnarray}
\epsilon_{\rm b}&\leq& \inf_{\delta\in(E_{\rm tol},1/2)}
\Bigg\{\left[1-\exp\left(\frac{(\delta{}N_{\rm tol} -\lfloor{}E_{\rm tol}N_{\rm tol}\rfloor)^2}{1-N_{\rm tol}}\right)\right]
2^{1-[1-h(\delta)]N_{\rm tol}} \nonumber \\
&&\quad
+2\exp\left(\frac{(\delta{}N_{\rm tol} -\lfloor{}E_{\rm tol}N_{\rm tol}\rfloor)^2}{1-N_{\rm tol}}\right)
\Bigg\}
\left[1+\sum_{k=1}^{\lfloor E_{\rm tol}N_{\rm tol}\rfloor}(2^k-1){N_{\rm tol} \choose k}\right].
\end{eqnarray}
After composing the errors related to the estimation of the parameters
$n_{\rm rect}$ and $n_{\rm diag}$ we obtain
Eq.~(\ref{main}).
\end{proof}

In Tab.~1 in the paper we have that $N_{\rm tol}=107$, $E_{\rm tol}=1.5\%$, and we select
$\varepsilon_{\rm rect}=\varepsilon_{\rm diag}=0.21\times10^{-2}$. Using
Eq.~(\ref{main}) we obtain, therefore, that $\epsilon_{\rm b}\leq 0.0568$, where the parameter
$\delta$ that minimises Eq.~(\ref{main}) is $\delta=0.2953$.
Since the protocol is perfectly concealing (see next section), this implies that
the committed bits are $\epsilon$-secure, with
$\epsilon\leq 0.0568$.

Let us remark that the upper bound given by Eq.~(\ref{main}) may not be tight, specially when the 
number of errors increases. However, since the error rate of our experiment is very low, this bound is 
enough for our purposes and we use it for simplicity. One way to improve this result would be to find a tighter upper 
bound for the l.h.s. of Eq.~(\ref{vir}). If, moreover, this bound is written in terms of the min-entropy $H_{\rm min}(S_{B_{\rm diag}}|A_1)$ then
all the security arguments used above could be applied directly.
In reality, therefore, the total cheating probability in the experiment may be significantly lower than $0.0568$. 

\subsection{Security against a dishonest Bob:}

Clearly,
if the probability that Alice detects Bob's signals is independent of the measurement basis selected,
the bit commitment protocol implemented is perfectly concealing. This is so because
Alice only informs Bob about which signals she
has actually detected and her communication with agents $A_0$ and $A_1$ is encoded with the one-time-pad (OTP)~\cite{vernam}.
This means that the probability that a dishonest Bob guesses Alice's committed bit correctly is $1/2$.

It is therefore essential for any experimental realisation
of the protocol to guarantee that Alice's detection probability
is independent of her measurement choice.
To illustrate this point, below we discuss briefly
some potential cheating strategies that a dishonest Bob may try to implement to
obtain the committed bit. They exploit different imperfections of Alice's
threshold detectors that result in a detection probability that depends
on Alice's basis selection.

\noindent{\it Exploiting double clicks:} Due to the background noise ({\it i.e.}, the dark counts of the
detectors together with other possible background contributions)
Alice may occasionally observe
a simultaneous click in her two detectors. Similar to the situation in quantum key distribution, double clicks should not be discarded by Alice but
they should be assigned to a random click, as we do in our experiment. Otherwise, a dishonest Bob may exploit double clicks to
obtain the committed bit. For instance, he could send Alice a very strong pulse in say horizontal
polarisation. Clearly, if Alice uses the rectilinear basis to measure the incoming pulse, she will observe a click in
the detector associated to horizontal polarisation. However, if she uses the diagonal basis,
she will observe a double click. If double clicks are discarded, then Bob will learn the committed bit
when Alice informs him about which pulses she detected.

\noindent{\it Exploiting the dead-time of Alice's detectors:} Similar to the previous case, a dishonest Bob
may also exploit
the dead-time of Alice's detectors to produce, or not to produce, a click depending on the measurement basis.
For instance, Bob may send Alice two strong pulses prepared in say horizontal and vertical polarisation, respectively,
and separated by a time interval less than the dead-time. Then, if Alice uses the rectilinear basis, both signals will produce a
click. However, if she uses the diagonal basis, the first signal generates a double click, while
the second signal remains undetected due to the dead-time
of the detectors. As above, when Alice informs Bob about which signals she detected
he learns the committed bit.

Even if Alice only accepts clicks which are separated
by a time interval greater than the dead-time of her detectors, a dishonest Bob can obtain the committed bit.
For instance, Bob may send Alice three consecutive strong pulses in the time instants $t$, $t+t_{\rm dead}/2$ and
slightly after $t+t_{\rm dead}$, where $t_{\rm dead}$ represents the dead-time of the detectors. Moreover, suppose that the
first signal is prepared in horizontal polarisation, while the second and the third signals are prepared in vertical polarisation.
Then, if Alice uses the rectilinear basis, she will observe a click in the first two instants. And she will
report Bob a detected event only in the first instant (since the first two instants are separated by a time interval
smaller than $t_{\rm dead}$ and, therefore, she discards the second click). However, if she uses the diagonal basis, she will observe a double click
in both the first and last instant. The second signal is never detected due
to the dead-time of the detectors. Again, the information about Alice's detected events
(in particular, whether or not the third signal is detected) reveals Bob
the committed bit.

To avoid this type of attacks and guarantee that the detection probability of Alice is independent
of her measurement choice, Alice needs to ensure that her measurement results originate from events
where both detectors were active. One simple solution to this problem is to actively control the dead-time. That is,
every time Alice observes a click in any of her detectors, she disable both detectors for a
time period equal to the dead-time. Alternatively, Alice may also post-select only those clicks that happened
after a time period of at least $2t_{\rm dead}$ without seeing any click. This last condition guarantees that
the post-selected clicks occurred when both detectors were active.
Indeed, this is the solution that we implemented in the experiment, where
the post-selection of data is performed in real time in an FPGA.

\section{Estimation of the parameters $n_{\rm rect}$, $n_{\rm diag}$, $e_{\rm rect}$ and $e_{\rm diag}$}\label{estimation}

In this Appendix we show how to estimate the foregoing parameters, which are used by Bob in the verification step of the protocol
to decide whether or not he accepts Alice's commitment.

We begin by introducing some notations. Let
$p_{\rm multi|rect}$ ($p_{\rm multi|diag}$) denote the conditional
probability that Bob sends Alice a signal containing two or more photons, given that he selected the rectilinear (diagonal) basis.
In the bit commitment protocol considered,
Bob sends Alice phase-randomised weak coherent pulses of intensity $\mu$. This means that
$p_{\rm multi|rect}$ and $p_{\rm multi|diag}$ satisfy
\begin{equation}\label{celta}
p_{\rm multi|rect}=p_{\rm multi|diag}\equiv
p_{\rm multi}\leq 1-\left[1+\mu\left(1+\delta\right)\right]e^{-\mu\left(1+\delta\right)},
\end{equation}
where the parameter $\delta$ denotes an upper bound for the intensity fluctuations of the laser diode.

Let $N_{\rm rect}$ ($N_{\rm diag}$) be the total number of signals sent by Bob in the rectilinear (diagonal) basis.
And, let $N_{\rm multi|rect}$ ($N_{\rm multi|diag}$) represent the total number of multi-photon signals sent by Bob
when he selects the
rectilinear (diagonal) basis.
The parameters $N_{\rm multi|rect}$ and $N_{\rm multi|diag}$ are not directly observed in the experiment but they can be
estimated.
Using Chernoff-Hoeffding inequality for i.i.d. random variables~\cite{cher,hoeff}, we have that
\begin{eqnarray}
N_{\rm multi|rect}\leq{}\lceil{}N_{\rm rect}(p_{\rm multi}+\delta_{\rm multi|rect})\rceil,
\end{eqnarray}
except with error probability $\varepsilon_{\rm rect}$ given by
\begin{equation}\label{eq1}
\varepsilon_{\rm rect}=e^{-D(p_{\rm multi}+\delta_{\rm multi|rect}||p_{\rm multi})N_{\rm rect}}.
\end{equation}
Here, $D(x||y)=x\ln{(x/y)}+(1-x)\ln{[(1-x)/(1-y)]}$
is the Kullback-Leibler divergence between Bernoulli distributed
random variables~\cite{kull}.
Similarly, we obtain that
$N_{\rm multi|diag}\leq{}\lceil{}N_{\rm diag}(p_{\rm multi}+\delta_{\rm multi|diag})\rceil$,
except with error probability $\varepsilon_{\rm diag}$ given by
\begin{equation}\label{eq2}
\varepsilon_{\rm diag}=e^{-D(p_{\rm multi}+\delta_{\rm multi|diag}||p_{\rm multi})N_{\rm diag}}.
\end{equation}

Finally, let $N_{\rm detect|rect}$ ($N_{\rm detect|diag}$) denote the total number of signals
declared as detected by Alice when Bob selected the rectilinear (diagonal) basis. Combining the results above,
we have that
\begin{eqnarray}\label{bound1}
n_{\rm rect}&\geq& N_{\rm detect|rect}-\lceil{}N_{\rm rect}(p_{\rm multi}+\delta_{\rm multi|rect})\rceil, \nonumber \\
n_{\rm diag}&\geq& N_{\rm detect|diag}-\lceil{}N_{\rm diag}(p_{\rm multi}+\delta_{\rm multi|diag})\rceil,
\end{eqnarray}
expect with error probability $\varepsilon\leq{}\varepsilon_{\rm rect}+\varepsilon_{\rm diag}$.
To derive Eq.~(\ref{bound1})
we have assumed the worse case scenario where all multi-photon signals sent by Bob are actually declared as detected by Alice.
The parameters $N_{\rm detect|rect}$, $ N_{\rm detect|diag}$, $N_{\rm rect}$ and $N_{\rm diag}$ are
observed in the experiment,
the probabilities
$p_{\rm multi|rect}$ and $p_{\rm multi|diag}$ are fixed by Bob's state preparation process, and the terms $\delta_{\rm multi|rect}$ and $\delta_{\rm multi|diag}$
can be obtained using Eqs.~(\ref{eq1})-(\ref{eq2}) for any given value of the tolerated error probabilities
$\varepsilon_{\rm rect}$ and $\varepsilon_{\rm diag}$. These quantities are shown
in Tab.~\ref{Tab:2838} for the experiment reported in the paper.

The calculation of $n_{\rm e,{\rm rect}}$ and $n_{\rm e,{\rm diag}}$ is straightforward. We 
consider a worse case scenario where all the errors observed are assumed to affect only
the single photon signals sent by Bob. That is, $n_{\rm e,{\rm rect}}$ ($n_{\rm e,{\rm diag}}$)
is directly given by the total number of errors in the rectilinear (diagonal) basis. 

In the experiment
reported in the paper $\mu=0.183$ and $\delta\leq0.1$. According to Eq.~(\ref{celta}), this means that $p_{\rm multi}\leq0.0177$.

%%%%%%%%%%%%%%%%%%%%%%

\begin{table}
\centering
\begin{tabular}{c|cccc}
\hline
\hline
 & $N_{\rm detect|rect}$ & $N_{\rm detect|diag}$ & $n_{\rm rect}$ & $n_{\rm diag}$  \\
\hline
Exp $1$   & $189$ & $193$ & $112$ & $116$  \\
Exp $2$  & $196$ & $197$ & $119$ & $120$  \\
Exp $3$    & $203$ & $184$ & $126$ & $107$  \\
Exp $4$   & $195$ & $205$ & $118$ & $128$ \\
Exp $5$  & $192$ & $192$ & $115$ & $115$  \\
Exp $6$   & $186$ & $199$ & $109$ & $122$ \\
Exp $7$  & $186$ & $218$ & $109$ & $141$  \\
Exp $8$  & $196$ & $227$ & $119$ & $150$  \\
\hline
\hline
\end{tabular}
\caption{Parameters $n_{\rm rect}$ and $n_{\rm diag}$ given by Eq.~(\ref{bound1})
when Bob sends Alice two parallel sequences of $2838$ pulses. They correspond
to the results reported in Tab.~1 in the paper.
The quantities $N_{\rm rect}=N_{\rm diag}=2838$, $p_{\rm multi}\leq0.0177$
and
$\varepsilon_{\rm rect}=\varepsilon_{\rm diag}=0.21\times10^{-2}$.
This implies
 $\delta_{\rm multi|rect}=\delta_{\rm multi|diag}=0.00937$.}
\label{Tab:2838}
\end{table}

\section{Estimation of the parameter $t_{\rm commit}$}

In this Appendix we show how to estimate an upper bound for $t_{\rm commit}$, {\it i.e.},
the latest time instant where Alice could have made her commitment, given that Bob accepted the revealed bit.

\begin{figure}
\begin{center}
 \includegraphics[scale=0.5]{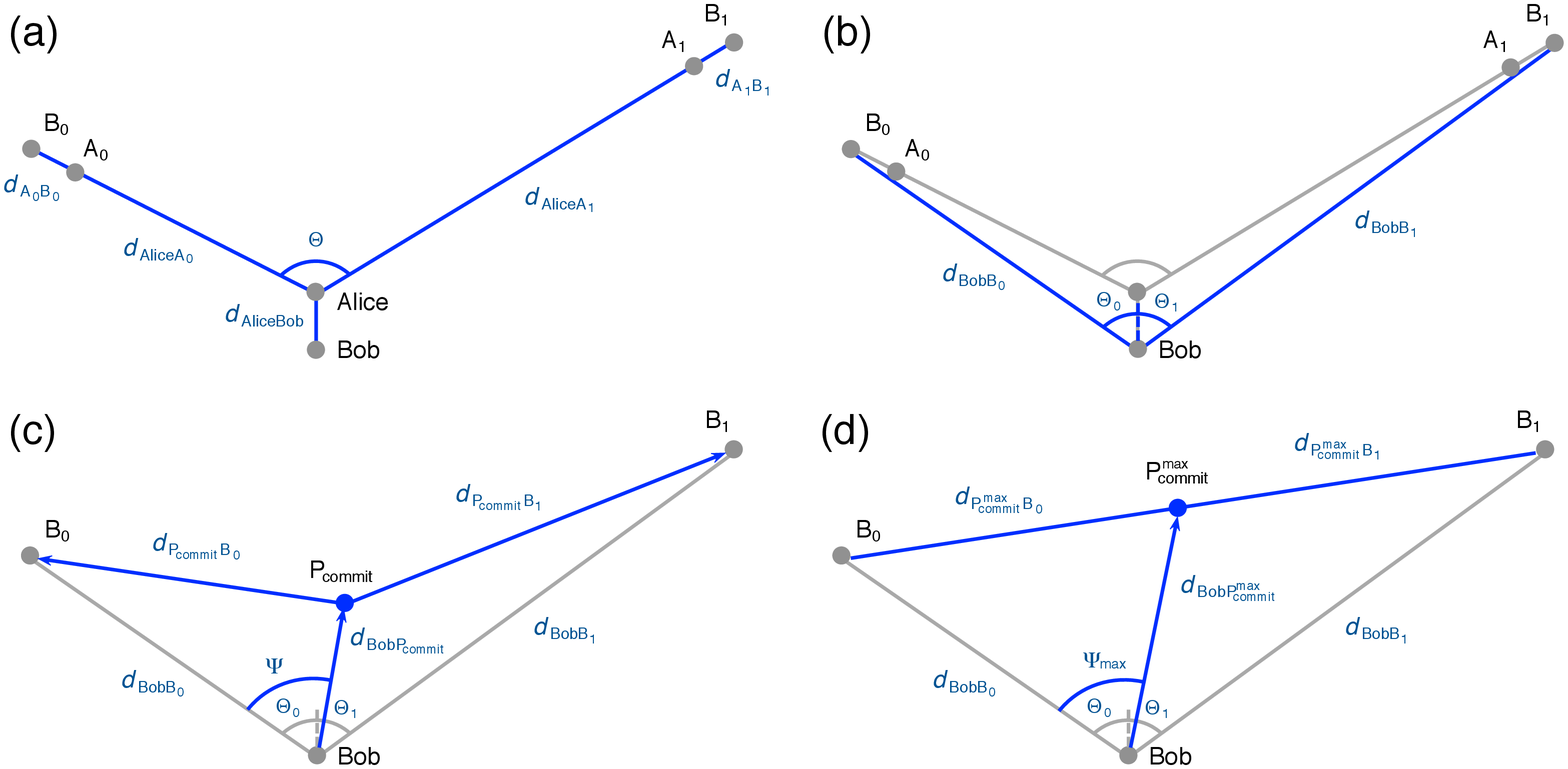}
 \end{center}
 \caption{(a) Geographical distribution
of the different parties involved in the protocol.
(b) We consider that
a dishonest Alice has access to all the $N$ signals that Bob sends her in one run of the protocol already at the time instant $t_0$
and in a space point
infinitely close to Bob.
(c) A dishonest Alice may store all the signals sent by Bob in a quantum memory and move the memory with her at the speed
of light to a certain
space point $P_{\rm commit}$ favourable for her commitment.
(d) The distance $d_{{\rm Bob}P_{\rm commit}}$
is always less or equal than a certain maximum distance $d_{{\rm Bob}P_{\rm commit}^{\rm max}}$.
If Alice has time to arrive at the space point $P_{\rm commit}^{\rm max}$ (in the line connecting the agents
$B_0$ and $B_1$), it is better for her (in order to increase the value of $t_{\rm commit}$) to stay at
$P_{\rm commit}^{\rm max}$ and wait there, rather than moving somewhere else.
\label{configuration}}
\end{figure}

Our starting point is the geographical distribution of the different parties involved in the protocol. It is illustrated in
Fig.~\ref{configuration}a.
Three important observables in the experiment are the time instants $t_0$, $t_{B_0}$ and $t_{B_1}$.
The first one represents the time instant where Bob sends Alice his first signal, while
$t_{B_0}$ ($t_{B_1}$) is the time instant where agent $B_0$ ($B_1$) receive the last
signal from agent $A_0$ ($A_1$). In order to find an upper bound for $t_{\rm commit}$ we will
assume, for simplicity, a very conservative scenario in which Alice has
access (at already the time instant $t_0$ and in a space point
infinitely close to Bob) to all the $N$ signals that Bob sends her in one run of the protocol.
Also, we will consider that $B_0$ and $B_1$ receive the signals from
$A_0$ and $A_1$ altogether at time instants $t_{B_0}$ and $t_{B_1}$, respectively.
This is illustrated in Fig.~\ref{configuration}b.

In practice, a dishonest Alice has some time to make her commitment, and typically she wants to commit as late as
possible. For this, she can store all the signals sent by Bob in a quantum memory and move the memory with her at the speed
of light to a certain
space point favourable for her commitment. This is illustrated in Fig.~\ref{configuration}c,
where $P_{\rm commit}$ denotes the
space point selected by Alice to make her commitment. This commitment point needs to satisfy the
following two conditions
\begin{eqnarray}\label{q1}
t_{B_0}&\geq&t_0+\frac{d_{{\rm Bob}P_{\rm commit}}+d_{P_{\rm commit}B_0}}{c}, \nonumber \\
t_{B_1}&\geq&t_0+\frac{d_{{\rm Bob}P_{\rm commit}}+d_{P_{\rm commit}B_1}}{c},
\end{eqnarray}
where $c$ denotes the speed of light in vacuum.
That is, the
total time needed to move the signals sent by Bob at $t_0$ through the paths $d_{{\rm Bob}P_{\rm commit}}$ and $d_{P_{\rm commit}B_0}$
($d_{{\rm Bob}P_{\rm commit}}$ and $d_{P_{\rm commit}B_1}$) should be less or equal to
$t_{B_0}$ ($t_{B_1}$).

The parameters $d_{P_{\rm commit}B_0}$ and $d_{P_{\rm commit}B_1}$ can be expressed as a function of the distance $d_{{\rm Bob}P_{\rm commit}}$ and the angle
of $B_0$-Bob-$P_{\rm commit}$, which we denote as $\Psi$, as
follows
\begin{eqnarray}\label{q2}
d_{P_{\rm commit}B_0}&=&\sqrt{d_{{\rm Bob}B_0}^2+d_{{\rm Bob}P_{\rm commit}}^2-2d_{{\rm Bob}B_0}d_{{\rm Bob}P_{\rm commit}}\cos{\Psi}}, \nonumber \\
d_{P_{\rm commit}B_1}&=&\sqrt{d_{{\rm Bob}B_1}^2+d_{{\rm Bob}P_{\rm commit}}^2-2d_{{\rm Bob}B_1}d_{{\rm Bob}P_{\rm commit}}\cos{(\theta_0+\theta_1-\Psi)}},
\end{eqnarray}
where
the distances $d_{{\rm Bob}B_i}$, together with the angles $\theta_i$ (see Fig.~\ref{configuration}b), with $i=0,1$, are given by
\begin{eqnarray}\label{q6}
d_{{\rm Bob}B_i}&=&\sqrt{d_{\rm AliceBob}^2+(d_{{\rm Alice}A_i}+d_{A_iB_i})^2+2d_{\rm AliceBob}(d_{{\rm Alice}A_i}+d_{A_iB_i})\cos{(\theta/2)}}, \nonumber \\
\theta_i&=&\arccos{\left\{\left[d_{\rm AliceBob}+(d_{{\rm Alice}A_i}+d_{A_iB_i})\cos{(\theta/2)}\right]/d_{{\rm Bob}B_i}\right\}},
\end{eqnarray}
where $\theta$ is the angle of $A_0$-Alice-$A_1$.

Combining these expressions with Eq.~(\ref{q1}) we have that
\begin{eqnarray}\label{qq1}
c(t_{B_0}-t_0)&\geq&d_{{\rm Bob}P_{\rm commit}}+\sqrt{d_{{\rm Bob}B_0}^2+d_{{\rm Bob}P_{\rm commit}}^2-2d_{{\rm Bob}B_0}d_{{\rm Bob}P_{\rm commit}}\cos{\Psi}}, \\
c(t_{B_1}-t_0)&\geq&d_{{\rm Bob}P_{\rm commit}}
+\sqrt{d_{{\rm Bob}B_1}^2+d_{{\rm Bob}P_{\rm commit}}^2-2d_{{\rm Bob}B_1}d_{{\rm Bob}P_{\rm commit}}\cos{(\theta_0+\theta_1-\Psi)}}. \nonumber
\end{eqnarray}

Also, the distance $d_{{\rm Bob}P_{\rm commit}}$
is always less or equal than $d_{{\rm Bob}P_{\rm commit}^{\rm max}}$ (see Fig.~\ref{configuration}d).
If Alice has time to arrive at the space point $P_{\rm commit}^{\rm max}$, it is better for her (in order to increase the value of $t_{\rm commit}$) to stay at
$P_{\rm commit}^{\rm max}$ and wait there, rather than continuing to move somewhere else. We have, in particular, that
\begin{eqnarray}\label{ffw}
d_{P_{\rm commit}^{\rm max}B_0}&=&qd_{B_0B_1}, \nonumber \\
d_{P_{\rm commit}^{\rm max}B_1}&=&(1-q)d_{B_0B_1},
\end{eqnarray}
with the distance $d_{B_0B_1}$ given by
\begin{equation}
d_{B_0B_1}=\sqrt{d_{{\rm Bob}B_0}^2+d_{{\rm Bob}B_1}^2-2d_{{\rm Bob}B_0}d_{{\rm Bob}B_1}\cos{(\theta_0+\theta_1)}},
\end{equation}
and where the term $q$ has the form
\begin{equation}\label{eq_q}
q=\frac{1}{2}\left[1-\frac{c}{d_{B_0B_1}}(t_{B_1}-t_{B_0})\right].
\end{equation}
This last result comes from the following. After the commitment, we have that (in the worse case scenario)
$d_{P_{\rm commit}^{\rm max}B_1}/c-d_{P_{\rm commit}^{\rm max}B_0}/c=t_{B_1}-t_{B_0}$. That is,
 using Eq.~(\ref{ffw}) we obtain $(1-2q)d_{B_0B_1}=c(t_{B_1}-t_{B_0})$, which is
 equivalent to Eq.~(\ref{eq_q}).

After some simple calculations, we obtain that $d_{{\rm Bob}P_{\rm commit}^{\rm max}}$ is given by
\begin{equation}
d_{{\rm Bob}P_{\rm commit}^{\rm max}}=\sqrt{\xi^2+(d_{{\rm Bob}B_1}-\delta)^2},
\end{equation}
 where the parameters $\xi$ and $\delta$ are given by, respectively,
 \begin{eqnarray}
 \xi&=&(1-q)d_{{\rm Bob}B_0}\sin{(\theta_0+\theta_1)}, \nonumber \\
 \delta&=&(1-q)d_{B_0B_1}\cos{(\beta)},
 \end{eqnarray}
 and where the angle $\beta$ has the form $\beta=\arcsin{[(d_{{\rm Bob}B_0}/d_{B_0B_1})\sin{(\theta_0+\theta_1)}]}$.

Based on the foregoing, we find that $d_{{\rm Bob}P_{\rm commit}}$ in Fig.~\ref{configuration}c can be obtained solving the
following constrained optimisation problem
\begin{eqnarray}\label{q5}
&&\max_{\Psi}\quad d_{{\rm Bob}P_{\rm commit}} \nonumber \\
&&{\rm s.t.}\quad 0\leq{}\Psi\leq \theta_0+\theta_1\nonumber \\
&&d_{{\rm Bob}P_{\rm commit}}+
\sqrt{d_{{\rm Bob}B_0}^2+d_{{\rm Bob}P_{\rm commit}}^2-2d_{{\rm Bob}B_0}d_{{\rm Bob}P_{\rm commit}}\cos{\Psi}}\leq{}c(t_{B_0}-t_0), \nonumber \\
&&d_{{\rm Bob}P_{\rm commit}}
+\sqrt{d_{{\rm Bob}B_1}^2+d_{{\rm Bob}P_{\rm commit}}^2-2d_{{\rm Bob}B_1}d_{{\rm Bob}P_{\rm commit}}\cos{(\theta_0+\theta_1-\Psi)}}\leq{}c(t_{B_1}-t_0),  \nonumber \\
&&0\leq{}d_{{\rm Bob}P_{\rm commit}}\leq{}\min\left[c(t_{B_0}-t_0),c(t_{B_1}-t_0),d_{{\rm Bob}P_{\rm commit}^{\rm max}}\right].
\end{eqnarray}
Let $d_{{\rm Bob}P_{\rm commit}}^{\rm sol}$ denote the solution to this optimisation problem.
This solution can be obtained using either analytical or numerical methods. Then we need to consider two cases.
If $d_{{\rm Bob}P_{\rm commit}}^{\rm sol}=d_{{\rm Bob}P_{\rm commit}^{\rm max}}$, this means that Alice has time to arrive at
$P_{\rm commit}^{\rm max}$
and wait there before she makes her commitment. That is,
combining Eqs.~(\ref{ffw})-(\ref{eq_q}) with the fact that
\begin{eqnarray}\label{neweq}
t_{B_0}&\geq&t_0+t_{\rm commit}+\frac{d_{P_{\rm commit}B_0}}{c}, \nonumber \\
t_{B_1}&\geq&t_0+t_{\rm commit}+\frac{d_{P_{\rm commit}B_1}}{c}.
\end{eqnarray}
we find that
$t_{\rm commit}$ is upper bounded by
\begin{eqnarray}\label{hgf}
t_{\rm commit}&\leq&t_{B_0}-t_0-\frac{qd_{B_0B1}}{c}=\frac{1}{2}\left(t_{B_0}+t_{B_1}-\frac{d_{B_0B_1}}{c}\right)-t_0 \nonumber \\
&\approx&\frac{1}{2}\left(t_{B_0}+t_{B_1}-\frac{d_{A_0A_1}}{c}\right)-t_0,
\end{eqnarray}
where in the last equality we used
$d_{A_0B_0},d_{A_1B_1}, d_{\rm AliceBob}, \ll d_{{\rm Alice}A_0},d_{{\rm Alice}A_1},d_{A_0A_1}$,
which means that $d_{B_0B_1}\approx{}d_{A_0A_1}$.
Eq.~(\ref{hgf})
corresponds to the typical situation where the total number of signals $N$
sent by Bob is large and thus also $t_{B_0}$ and $t_{B_1}$ are large.

The second case is when $d_{{\rm Bob}P_{\rm commit}}^{\rm sol}<d_{{\rm Bob}P_{\rm commit}^{\rm max}}$. In this situation
Alice has to make her commitment before she can arrive at
$P_{\rm commit}^{\rm max}$. This means that $t_{\rm commit}$ is directly given by
\begin{equation}
t_{\rm commit}=\frac{d_{{\rm Bob}P_{\rm commit}}^{\rm sol}}{c}.
\end{equation}
This is the minimum time needed to arrive at the point $P_{\rm commit}$ (see Fig.~\ref{configuration}c).

Finally, to verify that the commitment is not performed in certain space points like, for
instance, in the locations of $A_0$ and $A_1$, Bob needs to
confirm that the following two conditions are fulfilled:
$d_{{\rm Bob}A_0}+d_{A_0B_1}>c(t_{B_1}-t_0)$ and
$d_{{\rm Bob}A_1}+d_{A_1B_0}>c(t_{B_0}-t_0)$.
The first condition assures that there is no time for the signals to travel
from Bob to $A_0$, and from $A_0$ to $B_1$, before $t_{B_1}$ (and similarly for the
second condition).
When
$d_{A_0B_0},d_{A_1B_1}, d_{\rm AliceBob}, \ll d_{{\rm Alice}A_0},d_{{\rm Alice}A_1},d_{A_0A_1}$
these conditions reduce to the ones presented in the paper.

\end{document}